\documentclass[12pt,a4paper]{article}
\usepackage{}
\usepackage{amsmath,amssymb}
\usepackage[all]{xy}
 \usepackage{latexsym}
\usepackage{amsthm,a4wide}
\input diagxy

  \theoremstyle{plain}

  \theoremstyle{definition}

\begin{document}

\newcommand{\lam}{\lambda}
\newcommand{\da}{\downarrow}
\newcommand{\Da}{\Downarrow\!}
\newcommand{\D}{\Delta}
\newcommand{\ua}{\uparrow}
\newcommand{\ra}{\rightarrow}
\newcommand{\la}{\leftarrow}
\newcommand{\Lra}{\Longrightarrow}
\newcommand{\Lla}{\Longleftarrow}
\newcommand{\rat}{\!\rightarrowtail\!}
\newcommand{\up}{\upsilon}
\newcommand{\Up}{\Upsilon}
\newcommand{\ep}{\epsilon}
\newcommand{\ga}{\gamma}
\newcommand{\Ga}{\Gamma}
\newcommand{\Lam}{\Lambda}
\newcommand{\CF}{{\cal F}}
\newcommand{\CG}{{\cal G}}
\newcommand{\CH}{{\cal H}}
\newcommand{\CI}{{\cal I}}
\newcommand{\CB}{{\cal B}}
\newcommand{\CT}{{\cal T}}
\newcommand{\CS}{{\cal S}}
\newcommand{\CV}{{\cal V}}
\newcommand{\CP}{{\cal P}}
\newcommand{\CQ}{{\cal Q}}
\newcommand{\cu}{{\underline{\cup}}}
\newcommand{\ca}{{\underline{\cap}}}
\newcommand{\nb}{{\rm int}}
\newcommand{\Si}{\Sigma}
\newcommand{\si}{\sigma}
\newcommand{\Om}{\Omega}
\newcommand{\bm}{\bibitem}
\newcommand{\bv}{\bigvee}
\newcommand{\bw}{\bigwedge}
\newcommand{\lra}{\longrightarrow}
\newcommand{\tl}{\triangleleft}
\newcommand{\tr}{\triangleright}
\newcommand{\dda}{\downdownarrows}
\newcommand{\dia}{\diamondsuit}
\newcommand{\y}{{\bf y}}
\newcommand{\colim}{{\rm colim}}
\newcommand{\fR}{R^{\!\forall}}
\newcommand{\eR}{R_{\!\exists}}
\newcommand{\dR}{R^{\!\da}}
\newcommand{\uR}{R_{\!\ua}}
\newcommand{\cl}{{\rm cl}}
\title{{\Large Lagrangian Actions on Elliptical Solutions of 2-Body and 3-Body Problems with Fixed
Energies}\thanks{Supported partially by NSF of China}}
\author{ {\normalsize Ying Lv and Shiqing Zhang}\\
{\small College of Mathematics and Statistics, Southwestern
University, Chongqing 400715, P.R.China}\\ {\small Yangtze Center
of Mathematics, Sichuan University, Chengdu 610064, P.R.China}}
\date{}
\maketitle
\begin{abstract} Based on the works of Gordon ([4]) and Zhang-Zhou([8])) on the variational
minimizing properties for Keplerian orbits and Lagrangian
solutions of Newtonian 2-body and 3-body problems, we use the
constrained variational principle of Ambrosetti-Coti Zelati ([1])
to compute the Lagrangian actions on Keplerian and Lagrangian
elliptical solutions with fixed energies, we also find an
interesting relationship between period and energy for Lagrangian
elliptical solutions with Newtonian potentials.

\bf{2000MSC, 70G75, 70F07, 70F10.}
\end{abstract}
\parskip 1pt

\section{Introduction and  Main Results}
\hspace*{\parindent}

In [4], Gordon proved that the Keplerian orbits minimize the
Lagrangian action of the Keplerian 2-body problems with a fixed
period, in [8], Zhang-Zhou generalized the result of Gordon to
Newtonian 3-body problems and proved that the Lagrangian
elliptical orbits with equilateral configurations minimize the
Lagrangian action with a fixed period.

In this note, we try to generalize the above cases for the fixed
period to the fixed energy.

Consider Keplerian two-body problem with a fixed energy $h<0$:
\begin{equation}
\left\{\begin{array}{l}
\ddot{x}(t)+\nabla V(x)=0,\ \ \ \ \ \ \ x\in R^2\\
\frac{1}{2}|\dot{x}|^2+V(x)=h,
\end{array}\right.\label{(1.1)}
\end{equation}
where \begin{equation}
V(x)=\frac{-a}{|x|},\ \ \ \ a>0\label{(1.3)}
\end{equation}

Let $W^{1,2}(R/Z,R^2)$ denote the Sobolev space with period 1 and
the usual inner product and norm:
\begin{eqnarray}
&&<u,v>=\int^1_0(u\cdot v+\dot{u}\cdot \dot{v})dt\label{(1.4)}\\
&&||u||=\langle u, u\rangle^{1/2}\label{(1.5)}
\end{eqnarray}
For two-body problems with a fixed energy $h$,
Ambrozeth-CotiZelati ([1]) defined Lagrangian action:
\begin{equation}
f(u)=\frac{1}{2}\int^1_0|\dot{u}|^2dt\cdot\int^1_0(h-V(u))dt\label{(1.6)}
\end{equation}
and the following constrained manifold:
\begin{equation}
M_h=\left\{u\in W^{1,2}|u(t)\not\equiv 0\left|
\int^1_0(\frac{1}{2}V^{\prime}(u)u+V(u))dt=h\right.\right\}\label{(1.7)}
\end{equation}
and they proved that the critical point $\tilde{u}$ of $f(u)$ on
$M_h$ corresponds to a noncollision $T$-periodical solution
$\tilde{q}(t)=\tilde{u}(t/T)$ of the system (1) after a scaling
for the period $T$:
\begin{eqnarray}
\frac{1}{T^2}&=&\frac{\int^1_0V^{\prime}(\tilde{u})\cdot\tilde{u}dt}{\int^1_0|\dot{\tilde{u}}|^2dt}\nonumber\\
&=&\frac{\int^1_0(h-V(\tilde{u}))dt}{\frac{1}{2}\int^1_0|\dot{\tilde{u}}|^2dt}\label{(1.8)}
\end{eqnarray}
For $N$-body type problems, they also showed the similar
variational principle.

For two-body problem (1), we have the following Theorem:

{\bf Theorem 1.1}\ \ Let degu denote the winding number of the
loop $u$ respect to the origin, define
\begin{equation}
\Lambda_1=\{u\in M_h, {\rm deg}u\neq 0\}.\label{(9)}
\end{equation}
Then the global minimum of $f(u)$ on the closure
$\overline{\Lambda_1}$ exists and equals to $\frac{9}{16}\cdot
2^{-1/3}(\pi a)^2(-h)^{-1}$, and the minimizer $\tilde{u}(t)$ of
$f(u)$ on $\overline{\Lambda_1}$ are exactly corresponding to the
stright line collision solution $\tilde{x}(t)=u(t/T)$ or Keplerian
elliptical solution $x(t)=\tilde{u}(t/T)$ under a scaling
transform:
\begin{equation}
T=2\pi (-2h)^{-3/2}a\label{(1.9)}
\end{equation}
and $x(t)$ has energy $h$.

For Newtonian 3-body problems with a fixed enery $E$:
\begin{equation}
\left\{\begin{array}{l} m_i\ddot{q}_i=\frac{-\partial
V(q)}{\partial
q_i},\\
\frac{1}{2}\sum\limits^3_{i=1}m_i|\dot{q}_i|^2+V(q)=E,
\end{array}\right.\label{(1.10)}
\end{equation}
where \begin{equation} V(q)=-\sum\limits_{1\leq i<j\leq
3}\frac{m_im_j}{|q_i-q_j|}.\label{(13)}
\end{equation}

We define.
\begin{eqnarray}
&&F(u)=\frac{1}{2}\int^1_0\sum\limits^3_{i=1}m_i|\dot{u}_i|^2dt\cdot\int^1_0(E-V(u))dt,\label{(1.12)}\\
&&u\in\Lambda_2=\left\{\begin{array}{l} u=(u_1,u_2,u_3)|u_i\in
W^{1,2},\ \sum\limits^3_{i=1}m_iu_i=0,\\
{\rm deg}(u_i-u_j)\neq 0,\\
\int^1_0(V(u)+\frac{1}{2}V^{\prime}(u)u)dt=E
\end{array}\right\}\label{(1.13)}
\end{eqnarray}
Then we have

{\bf Theorem 1.2}\ \ The global minimizers of $f(u)$ on
$\Lambda_2$ are just the Lagrangian elliptical solutions and the
period for the elliptical orbits is
\begin{equation}
T=2\pi\cdot\left(\frac{\sum\limits_{1\leq i<j\leq
3}m_im_j}{-2E}\right)^{3/2}\label{(1.14)}
\end{equation}
and the Lagrangian action is
\begin{equation}
2^{-13/3}(3\pi )^2\left(\sum\limits_{1\leq i<j\leq
3}m_im_j\right)^3\cdot (-E)^{-1},\label{(1.15)}
\end{equation}

\section{The Proof of Theorem 1.1}

{\bf Lemma 2.1}(Newton [6])\ \ For Keplerian elliptical orbits of
two-body problem (1), the period $T$ and energy $h$ has the
following relationship:
\begin{equation}
T=2\pi (-2h)^{-3/2}a\label{(2.1)}
\end{equation}
{\bf Lemma 2.2}(Gordon [4])\ \ Let $\bar{\Lambda}$ be the
$W^{1,2}(R/TZ,R^2)$ completion of the following loop space with
period $T$:
\begin{equation}
\Lambda =\{x(t)\in C^{\infty}(R/TZ,R^2)|x(t)\neq 0, {\rm deg}x\neq
0\}\label{(2.2)}
\end{equation}
We define the Lagrangian action:
\begin{equation}
g(x)=\int^T_0\left(\frac{1}{2}|\dot{x}|^2+\frac{a}{|x|}\right)dt\label{(2.3)}
\end{equation}

Then the minimizers of $g(x)$ on $\bar{\Lambda}$ are the Keplerian
elliptical solutions or the straight line collision solution with
one leg, and the minimum is
\begin{equation}
(3\pi )\left(\frac{T}{2\pi}\right)^{1/3}\cdot
a^{2/3}=\frac{3}{2}(2\pi )^{2/3}a^{2/3}T^{1/3}.\label{(2.4)}
\end{equation}

{\bf Lemma 2.3}([3])\ \ Let $u(t)$ be a critical point of $f(u)$
on $\bar{\Lambda}$ and let $x(t)=u(t/T)$, then
\begin{eqnarray}
[4f(u)]^{1/2}&=&\int^T_0\left[\frac{1}{2}|\dot{x}|^2+\frac{1}{2}V^{\prime}(x)x\right]dt\nonumber\\
&=&\int^T_0\left[\frac{1}{2}|\dot{x}|^2+\frac{a}{2}\frac{1}{|x|}\right]dt\label{(2.5)}
\end{eqnarray}
Now we can prove Theorem 1.1:

By Lemmas 2.1-2.3, we have
\begin{eqnarray}
[4f(u)]^{1/2}&\geq&\frac{3}{2}(2\pi
)^{2/3}\left(\frac{a}{2}\right)^{2/3}T^{1/3}\nonumber\\
&=&\frac{3}{2}\pi^{2/3}a^{2/3}(2\pi
)^{1/3}(-2h)^{-1/2}a^{1/3}\label{(2.6)}\\
f(u)&\geq&\frac{9}{16}2^{-1/3}(\pi a)^2(-h)^{-1}\label{(2.7)}
\end{eqnarray}
and $f(u)$ attains the infimum if and only if the minimizers are
Keplerian elliptical orbits or the collision solution with one
leg.

\section{The Proof of Theorem 1.2}

{\bf Lemma 3.1}\ \ For a Lagrangian elliptical solution ([5])
$q=(q_1,q_2,q_3)$ with period $T$, the energy $E$ for masses
$m_1,m_2,m_3$ is
\begin{equation}
E=\left(-\frac{1}{2}\right)\left(\frac{T}{2\pi}\right)^{-2/3}\left(\sum\limits_{1\leq
i<j\leq 3}m_im_j\right).\label{(3.1)}
\end{equation}

{\bf Proof.}\ \ The Lagrangian solution ([5]) is
\begin{equation}
q(t)=x(t)(\alpha_1,\alpha_2,\alpha_3),\label{(3.2)}
\end{equation}
where
$|\alpha_1-\alpha_2|=|\alpha_2-\alpha_3|=|\alpha_3-\alpha_1|=1,
x(t)$ is the Keplerian elliptical orbit satisfying
\begin{equation}
\ddot{x}(t)=\frac{-ax(t)}{|x(t)|^3}\label{(3.3)}
\end{equation}
From (24),we have
\begin{eqnarray}
&&q_i(t)-q_j(t)=x(t)(\alpha_i-\alpha_j),\label{(3.4)}\\
&&\dot{q}_i(t)-\dot{q}_j(t)=\dot{x}(t)(\alpha_i-\alpha_j),\label{(3.5)}
\end{eqnarray}
\begin{eqnarray}
&&\frac{1}{2}|\dot{q}_i-\dot{q}_j|^2-\frac{M}{|q_i-q_j|},\nonumber\\
&=&\frac{1}{2}|\dot{x}|^2-\frac{M}{|x|}\triangleq h\label{(3.6)}
\end{eqnarray}
where $M=\sum\limits^3_{i=1}m_i$.

We notice that the energy for the Lagrangian elliptical solutions
is
\begin{eqnarray}
E&=&\frac{1}{2}\sum
m_i|\dot{q}_i|^2-\sum\limits_{i<j}\frac{m_im_j}{|q_i-q_j|}\nonumber\\
&=&\frac{1}{M}\sum\limits_{i<j}m_im_j\left[\frac{|\dot{q}_i-\dot{q}_j|^2}{2}-
\frac{M}{|q_i-q_j|}\right]\nonumber\\
&=&\frac{1}{M}\sum\limits_{i<j}m_im_jh\label{(3.7)}
\end{eqnarray}

For Keplerian orbits $(q_i-q_j)$, we use Lemma 2.1 to get
\begin{eqnarray}
&&T=2\pi (-2h)^{-3/2}\cdot M,\label{(3.8)}\\
&&\left(\frac{T}{2\pi M}\right)^{-2/3}=-2h\label{(3.9)}
\end{eqnarray}
Hence
\begin{eqnarray}
&&E=\left(\sum\limits_{i<j}m_im_j\right)\left(-\frac{1}{2}\right)\cdot\left(\frac{T}{2\pi
M}\right)^{-2/3},\label{(3.10)}\\
&&T=2\pi\left(\frac{\sum\limits_{i<j}m_im_j}{-2E}\right)^{3/2}\cdot
M\label{(3.11)}
\end{eqnarray}

{\bf Lemma 3.2}([7])\ \ Let $u=(u_1,u_2,u_3)$ be a critical point
of $F(u)$ on $\Lambda_2$, let $q(t)=u(t/T)$, then

\begin{eqnarray}
E&=&\frac{1}{2}\sum
m_i|\dot{q}_i|^2-\sum\limits_{i<j}\frac{m_im_j}{|q_i-q_j|}\nonumber\\
\label{(3.7)}
\end{eqnarray}
\begin{eqnarray}
[4F(u)]^{1/2}&=&\int^T_0\left[\frac{1}{2}\sum\limits^3_{i=1}m_i|\dot{q}_i|^2+\frac{1}{2}
V^{\prime}(q)\cdot q\right]dt\label{(3.12)}\\
&=&\int^T_0\left[\frac{1}{2}\sum\limits^3_{i=1}m_i|\dot{q}_i|^2-\frac{1}{2}V(q)\right]dt\label{(3.13)}
\end{eqnarray}
Similar to [8], we have
\begin{equation}
\sum\limits_im_i|\dot{q}_i|^2=\frac{1}{M}\sum\limits_{i<j}m_im_j|\dot{q}_i-\dot{q}_j|^2,\label{(3.14)}
\end{equation}
Hence
\begin{equation}
[4F(u)]^{1/2}=\int^T_0\frac{1}{M}\sum\limits_{i<j}m_im_j\left[\frac{1}{2}\left|\dot{q}_i-
\dot{q}_j\right|^2+\frac{M}{2}\frac{1}{|q_i-q_j|}\right]\label{(3.15)}
\end{equation}

By Gordon's Lemma 2.2,
\begin{equation}
[4F(u)]^{1/2}\geq\frac{1}{M}\left(\sum\limits_{i<j}m_im_j\right)\cdot\left[\frac{3}{2}
(2\pi
)^{2/3}\left(\frac{M}{2}\right)^{2/3}T^{1/3}\right]\label{(3.16)}
\end{equation}
and $[4F(u)]^{1/2}$ attains the infimum if and only if  for $1\leq
i\not= j\leq 3,$
\begin{equation}
\int^T_0\left[\frac{1}{2}|\dot{q}_i-\dot{q}_j|^2+\frac{M}{2}\frac{1}{|q_i-q_j|}\right]
dt=\frac{3}{2}(2\pi
)^{2/3}\left(\frac{M}{2}\right)^{2/3}T^{1/3}\label{(3.17)}
\end{equation}

Then similar to the proof in Zhang-Zhou [8],the equations (39)
hold if and only if $q=(q_1,q_2,q_3)$ is a Lagrangian elliptical
solution,so we know the minimizers of $[4F(u)]^{1/2}$ correspond
to Lagrangian elliptical solutions after a scaling.

By Lemma 3.1, we have
\begin{eqnarray}
[4F(u)]^{1/2}&\geq&\frac{3}{2}(2\pi
)^{2/3}\left(\frac{1}{2}\right)^{2/3}M^{-1/3}\left(\sum\limits_{i<j}m_im_j\right)\cdot
(2\pi
)^{1/3}\cdot\left(\frac{\sum\limits_{i<j}m_im_j}{-2E}\right)^{1/2}\cdot
M^{1/3}\nonumber\\
&=&\frac{3}{2}(2\pi
)(\frac{1}{2})(\frac{1}{2})^{\frac{1}{6}}\cdot\left(\sum
m_im_j\right)^{3/2}\cdot (-E)^{-1/2}\label{(3.18)}\\
F(u)&\geq&2^{-\frac{13}{3}}\cdot (3\pi
)^2\cdot\left(\sum\limits_{i<j}m_im_j\right)^3\cdot (-E)^{-1}
\label{(3.19)}
\end{eqnarray}

From the above proof ,we know that $F(u)$ attain the infimum on
$\Lambda_2$ if and only if the minimizers correspond Lagrangian
elliptical solutions after a scaling and the Lagrangian action on
Lagrangian elliptical solutions has the value in (15).

\end{document}